\documentclass[10pt,conference]{IEEEtran}
\usepackage{graphicx} 
\IEEEoverridecommandlockouts
\usepackage{cite}
\usepackage{amsmath,amssymb,amsfonts}
\usepackage{algorithmic}
\usepackage{graphicx}
\usepackage{textcomp}
\usepackage{xcolor}
\def\BibTeX{{\rm B\kern-.05em{\sc i\kern-.025em b}\kern-.08em
    T\kern-.1667em\lower.7ex\hbox{E}\kern-.125emX}}

\usepackage{CJKutf8}
\usepackage{amsfonts}
\usepackage[colorlinks=true,linkcolor=black,citecolor=black,urlcolor=black]{hyperref}
\usepackage{url}
\usepackage{booktabs}
\usepackage{multirow} 
\usepackage{colortbl}

\usepackage{graphicx}
\usepackage{subcaption} 
\usepackage{caption}

\usepackage{comment}
\usepackage{fancyhdr}
\usepackage{relsize} 
\newcommand{\smallbf}[1]{{\smaller{\textbf{#1}}}}

\usepackage{tcolorbox}
\newcommand{\answer}[2] {
    \begin{tcolorbox}[boxrule=0.5pt,left=1pt,right=1pt,top=1pt,bottom=1pt]
        \textit{\smallbf{Answer to RQ#1:}} #2
    \end{tcolorbox}
}

\usepackage{ifthen}
\newboolean{hidecomments}
\setboolean{hidecomments}{false}
\ifthenelse{\boolean{hidecomments}}
{\newcommand{\nb}[2]{}}
{\newcommand{\nb}[2]{
    \fbox{\bfseries\sffamily\scriptsize#1}
    {\sf\small$\blacktriangleright$ 
      {#2} $\blacktriangleleft$}}} 

\newcommand\qaname{CoReQA}

\title{CoReQA: Uncovering Potentials of Language Models in Code Repository Question Answering}

\author{Anonymous Authors}
\author{
\IEEEauthorblockN{ \centering
    Jialiang Chen\IEEEauthorrefmark{2}\IEEEauthorrefmark{4}\IEEEauthorrefmark{5},
    Kaifa Zhao\IEEEauthorrefmark{3}\IEEEauthorrefmark{4}\IEEEauthorrefmark{5}, 
    Jie Liu\IEEEauthorrefmark{6}\IEEEauthorrefmark{1}, 
    Chao Peng\IEEEauthorrefmark{6},
    Jierui Liu\IEEEauthorrefmark{6},
    Hang Zhu\IEEEauthorrefmark{6},
     }
\IEEEauthorblockN{ \centering
        Pengfei Gao\IEEEauthorrefmark{6},
        Ping Yang\IEEEauthorrefmark{6},
        Shuiguang Deng \IEEEauthorrefmark{2}\IEEEauthorrefmark{1}
    }
    \thanks{\IEEEauthorrefmark{4} Equal contribution (co-first authors). Authors are listed alphabetically by last name.}
    \thanks{\IEEEauthorrefmark{5} The job was done when the author was internship at ByteDance.}
    \thanks{\IEEEauthorrefmark{1} The corresponding authors.}

    \IEEEauthorblockA{
        {
            \IEEEauthorrefmark{2} Zhejiang University, 
            \IEEEauthorrefmark{3} The Hong Kong Polytechnic University,
            \IEEEauthorrefmark{6} ByteDance
        }
    }

}

\date{ICSE'2025 Abs 26 July, DDL 2 Aug }

\begin{document}

\maketitle

\begin{abstract}

Large language models that enhance software development tasks, such as code generation, code completion, and code question answering (QA), have been extensively studied in both academia and the industry.
The models are integrated into popular intelligent IDEs like JetBrains and Cursor. 
Current benchmarks for evaluating models' code comprehension capabilities primarily focus on code generation or completion, often neglecting QA, which is a crucial aspect of understanding code.
Existing code QA benchmarks are derived from code comments with predefined patterns (e.g., CodeQA) or focus on specific domains, such as education (e.g., CS1QA).
These benchmarks fail to capture the real-world complexity of software engineering and user requirements for understanding code repositories.

To address this gap, we introduce \qaname, 
a benchmark 
for \textbf{Co}de \textbf{Re}pository-level question answering, constructed from GitHub issues and 
comments from 176 popular repositories across four programming languages. 
Repository-level QA requires models to retrieve the relevant content from the repository and generate answers for questions.
Since questions and answers may include both natural language and code snippets, traditional evaluation metrics such as BLEU are inadequate for assessing repository-level QA performance.
Thus, we provide an LLM-as-a-judge framework to evaluate QA performance from five aspects. 
Based on \qaname, we evaluate the performance of three baselines, including two short-context models using generic retrieval strategies and one long-context model that utilizes the entire repository context.
Evaluation results show that state-of-the-art proprietary and long-context models struggle to address repository-level questions effectively.
Our analysis highlights the limitations of language models in assisting developers in understanding repositories and suggests future directions for improving repository comprehension systems through effective context retrieval methodologies.

\end{abstract}


\section{Introduction}
Large language models (LLMs)~\cite{deepseek2024, chatgpt2024, claude2024, gemini2024} demonstrate exceptional ability in understanding and processing natural language.
Recently, LLMs trained on extensive code datasets (Code-LLMs)~\cite{guo2024deepseek, cursor2024, nam2024using,bai2023qwen} show promising results in software engineering tasks.
The tasks include code completion~\cite{zhang2023repocoder,li2023cctest,nie2023learning}, code summarization~\cite{ahmed2022few}, bug fixing~\cite{sobania2023analysis, ni2020analyzing, drain2021generating}, and et.al. 

To alleviate burdens of the tasks in daily development, IDEs such as Cursor~\cite{cursor2024}, Copilot~\cite{GitHubCopilot}, and JetBrains~\cite{JetBrainsWebsite} integrate Code-LLMs into their features. 
The features include automatic code generation within code files, bug detection and repair suggestions in the terminal, and answering user questions in chat interfaces, among others.
Therefore, evaluating LLM capabilities is crucial, as it ensures that the models provide accurate, complete, and readable answers essential for software development. 
Rigorous evaluation helps identify the strengths and limitations of LLMs, guiding further improvements and ensuring that the models can effectively handle complex and real-world scenarios in software repositories.
However, existing evaluations predominantly focus on code generation~\cite{yu2024codereval, zhang2024codeagent, chen2021humaneval,ding2023crosscodeeval} or bug fixing~\cite{jimenez2023swe}, with limited attention given to code question answering.

While several recent evaluation studies introduce code QA datasets~\cite{sahu2024codequeries, liu2021codeqa, lee2022cs1qa}, the datasets primarily focus on simple yes-or-no question, and are limited method-level ~\cite{liu2021codeqa, bansal2021neural} or file-level~\cite{sahu2024codequeries} context.
The datasets do not reflect real-world scenarios that require understanding entire projects or clarifying inter-function dependencies, which are complex and large-scale in nature.
Responses in repository-level QA require natural language explanations accompanied by relevant code examples.
Existing repository-level benchmarks are limited to bug fixing~\cite{jimenez2023swe} and code generation~\cite{yu2024codereval} tasks.
The literature still lacks a repository-level code QA benchmark.

\begin{figure*}[t!]
    \centering
    \includegraphics[width=0.75\linewidth]{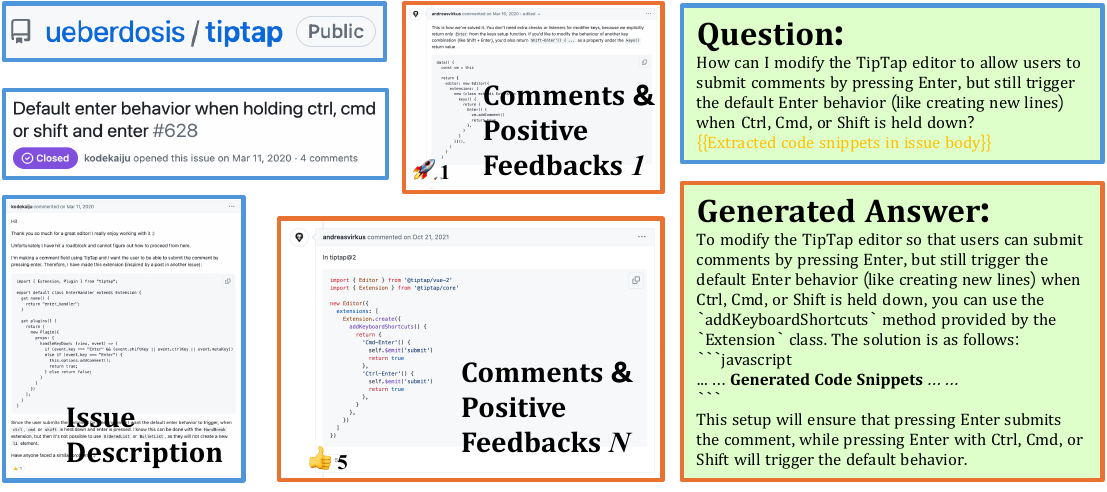}
    \caption{A question-answer pair from a sample GitHub issue and comments.}
    \label{fig:showcase}
\end{figure*}

To address this gap, we introduce \qaname, a benchmark designed for repository-level question answering. 
\qaname\ is crafted to reflect the complexity and diversity of real-world inquiries, facilitating a more precise evaluation of QA systems' capabilities within extensive code repositories.
To establish \qaname\ as a credible benchmark, we develop an GitHub issue-based QA generation framework. 
We collect real-world GitHub issues and curate a dataset of 1,563 QA pairs from 190 repositories across four programming languages.
We use LLMs to reformulate issue titles and descriptions into questions, making the issues more suitable for a question-answering context. 
To generate reference answers, we design a prompt and require language models to understand the semantics of the question, the raw issue content, and the relevant comments.
Fig.~\ref{fig:showcase} gives an example from \qaname, which is extracted from a closed GitHub issue in a popular repository and includes three positive comments.
GitHub issues reflect users' requirements for using the repository, while positive comments indicate agreement from other users. 
Our dataset construction framework ensures that demands represented are authentic and relevant to developers.


To evaluate the effectiveness of repository-level QA, we implement a comprehensive evaluation framework using the LLM-as-a-judge~\cite{NEURIPSllmjudge, es2023ragas}.
Our framework assesses the performance of models from two aspects: 
1) absolute quality evaluation that measures accuracy, completeness, relevance, and clarity, and 2) pairwise comparison evaluation for inter-model performance assessment.
We evaluate three state-of-the-art LLMs (GPT-4o~\cite{chatgpt2024}, Gemini-1.5~\cite{gemini2024}, and DeepSeek-V2~\cite{deepseek2024}) on our \qaname.
Addressing QA pairs in \qaname\ requires content from the repository. 
For instance, an issue may contain code snippets and pose questions, while the corresponding comments might include additional code snippets that address these questions, as illustrated in Figure~\ref{fig:showcase}.
Thus, we evaluate the baselines under three settings:
1) query the models without any contents,
2) query the models with general retrieval strategies, i.e, BM25~\cite{robertson2004simple},
and 3) provide the models the contents of the whole repository.
We use the third setting to evaluate long-context models on repositories where the text length fits within the model's input token limit, such as 10 million tokens for Gemini-1.5.



%
Experiments demonstrate that models achieve limited success on \qaname\ without additional content, with average scores of 6.37, 5.70, 7.33, and 8.09 out of 10 in accuracy, completeness, relevance, and clarity, respectively. 
When relevant context is provided, scores improve to 6.40, 5.74, 7.36, and 8.11 in accuracy, completeness, relevance, and clarity.
The scores improved slightly with retrieved relevant context, demonstrating the benchmark's challenge and the need for enhanced retrieval strategies to improve model performance.
%
Additionally, even when providing the entire repository content to long-context models, performance improves only slightly, and the metrics remain suboptimal.
Our results underscore the need for further research to develop effective content retrieval strategies.

Our contributions are summarized as follows: 
\begin{itemize}
\item We provide an automated framework to construct repository-level QA pairs and a novel evaluation infrastructure to assess the performance of generative models in addressing repository-level questions.

\item We introduce \qaname, a benchmark for repository question answering derived from GitHub repositories across various domains and programming languages. The questions are sourced from real-world user issues, with answers from corresponding discussions and comments.
%
%
%

\item We conduct a comprehensive evaluation and comparison on state-of-the-art LLMs on \qaname. 
Our evaluations include short-context LLMs with a general retrieval strategy and long-context LLMs for cross-file challenges.
Evaluation results highlight the ongoing difficulty of repository-level QA tasks for LLMs and the urgent need for precise context retrieval strategies.

\end{itemize}
%

\noindent\textbf{Roadmap.} 
The remainder of the paper is organized as follows: 
\S\ref{sec:background}  provides basic background on 
code comprehension and retrieval argument generation systems, and formulates repository-level QA task. 
%
%
\S\ref{sec:repoqa} introduces the framework of \qaname. 
\S\ref{sec:evaluation} presents the experimental settings and results of baseline models. 
\S\ref{sec:relatedwork} reviews related work.
\S\ref{sec:discussion} discusses threats to validity, ethical considerations, limitations, and future work.
\S\ref{sec:conclusion} concludes this work.

\section{Background}
\label{sec:background}
This section provides background on LLM-based code comprehension, presents prevalent strategies (e.g., retrieval-augmented generation and long-context learning) for solving repository-level code understanding tasks, and formally define the repository-level question answering task.

\begin{figure*}[]
    \centering
    \includegraphics[width=0.98\linewidth]{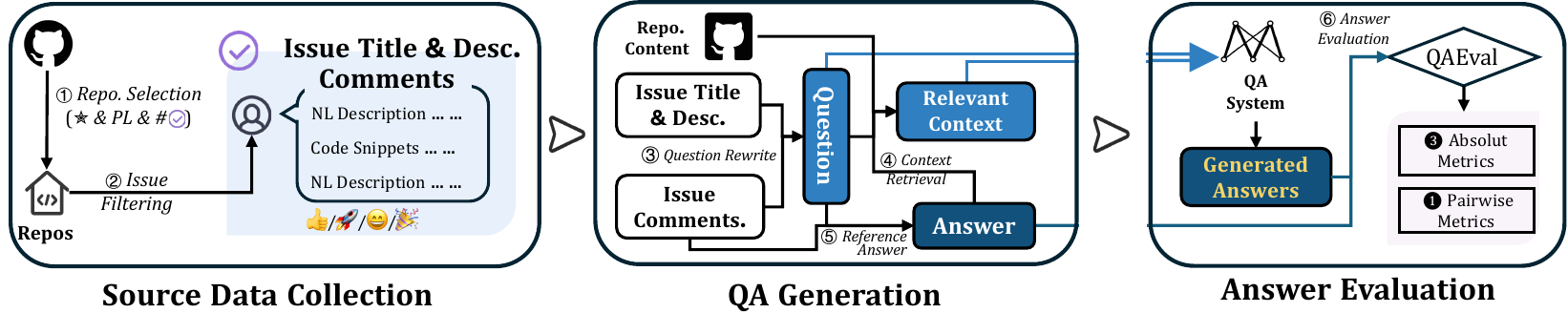}
    \caption{The overall pipeline of \qaname.}
    \label{fig:repoqa_fw}
\end{figure*}

\subsection{LLM-based code comprehension}
In the software development, 
code comprehension proves indispensable~\cite{ben2018neural,porkolab2018codecompass,bertolotti2023fold2vec} due to the iterative dependencies among code elements that collectively enable intricate functionalities. 
For instance, to automatically generate a code segment~\cite{zhang2024codeagent, luo2024repoagent}, developers need to understand the imported modules and discern between identically named functions belonging to different classes. 
Similarly, crafting an accurate pull request~\cite{jimenez2023swe} for a repository necessitates a clear grasp of requirements, alongside the aggregation of pertinent data, logs, and error messages.

Despite the advancements in large language models~\cite{chatgpt2024,gemini2024,deepseek2024,claude2024}, effectively tackling cross-file code comprehension remains a substantial hurdle. 
Interpreting the intent behind cross-file requirements is often inconsistent due to inherent ambiguity~\cite{9001577,pearce2023build}.
Additionally, retrieving relevant contents scattered across different files is challenging~\cite{es2023ragas, luo2024repoagent}.
Constructing the appropriate context to identify the core issue and develop a resolution is also a complex task.
The challenges underscore the urgent need for advancements in models and methods that can effectively navigate and understand the intricate interdependencies within code repositories.

\subsection{RAG and long-context learning}

Retrieval-augmented generation (RAG) enhances natural language processing (NLP) by integrating retrieval-based techniques with generative models.
RAG addresses the limitations of LLMs that often lack access to real-time and comprehensive external information. 
RAG framework combines a retriever component that fetches relevant content from a large corpus with a generator component that synthesizes this information into coherent responses~\cite{dong2022survey}.
The retriever identifies pertinent content based on the input query, while the generator uses the content to craft a detailed and accurate response~\cite{ren2023retrieve, li2024matching}. 
The combination of retriever and generator enhances the utility of LLMs in applications requiring up-to-date or specific knowledge.
In the domain of code repositories, the RAG framework retrieves relevant code snippets from various files and generates comprehensive explanations or solutions~\cite{luo2024repoagent, lee2022cs1qa}.
The RAG workflow involves processing a user query, retrieving pertinent documents, and generating a synthesized answer, ensuring that the response is both accurate and informative \cite{liu2021codeqa}. 
The RAG workflow enhances the accuracy of responses by integrating real-time and relevant information~\cite{jimenez2023swe, ding2023crosscodeeval, liu2021codeqa}. 
The approach leverages dynamic and external data that static models may miss. 

Long-context learning is an emerging solution~\cite{poli2023longcontext,dai2019transformer,zhang2024longcontext} for repository-level code comprehension, a methodology renowned for its capacity to analyze the entirety of a repository's content.
Nonetheless, this approach encounters limitations when repositories become excessively large, surpassing the feasible boundaries of what long-context models can efficiently process.

\subsection{Repository-level question answering task}
In this work, we focus on constructing a benchmark and setting up baselines for repository QA tasks. 
Focusing on a target code repository $\mathcal{R}$ and confronted with a repository-related question $q$, which comprises natural language, function or class signatures, or even code snippets, the task necessitates the production of a free-form textual answer $a$. 
The textual answer $a$ is versatile, encompassing the possibility of being a sentence, a phrase, or even a code snippet, contingent upon the context and requirements of the query. 
Indeed, distilling the answer $a$ from a single file would be overly simplistic and inadequate given that $q$ aims to probe or leverage insights about the entire repository  $\mathcal{R}$. 
Consequently, formulating a satisfactory response $a$ necessitates a deep comprehension of the repository's architecture alongside the retrieval and synthesis of pertinent information scattered throughout $\mathcal{R}$. 
Therefore, the repository-level question-answering task is a difficult challenge, which is exacerbated by the inherent scalability issues of large code bases and the complexity of long-term context understanding.
It is of immense significance and urgency to establish and disseminate a rigorous, representative benchmark for this domain.

\section{\qaname}
\label{sec:repoqa}

In this section, we introduce the construction process of the \qaname\, as illustrated in Fig.\ref{fig:repoqa_fw}.
The construction of the \qaname\ encompasses raw data collection, question-answer (QA) pair generation, and QA evaluation.

\subsection{\qaname\ construction}
\label{sec:construction}

\subsubsection{Raw Data Collection}
%

We design following pipelines to make 
\qaname\ authentic, reliable, and comprehensive.

\noindent \textbf{\textit{Repository Selection.}}
We start by selecting the top 500 GitHub repositories, ranked by the number of stars and having valid licenses\cite{huggingface_bigcode_2023}
 across four programming languages: Python, Java, Go, and TypeScript. 
Next, we filter out repositories where the number of closed issues and pull requests is either lower than 1,000 or higher than 50,000.
The selection ensures the repositories are not only popular but also reflective of active development communities in these prominent programming languages. 
We set a maximum threshold of 50k to reduce the analysis burden because every time we analyze the issue, we need to request GitHub API once.
We obtain 890 repositories.
%
%


To ensure that \qaname\ is capable of evaluating long-context models, token lengths for each repository are estimated using OpenAI's tokenizer calculation method~\cite{openai-tokenizer}:
the token length is approximated by dividing the number of characters by four.
We then select 50 repositories with token lengths greater than 200K and 5 with less than 200K~\cite{gemini2024, claude2024} for each programming language.
To this end, we obtain 218 candidate repositories with 1,623,624 issues.



\begin{figure}[t!]
    \centering
    \includegraphics[width=0.95\linewidth]{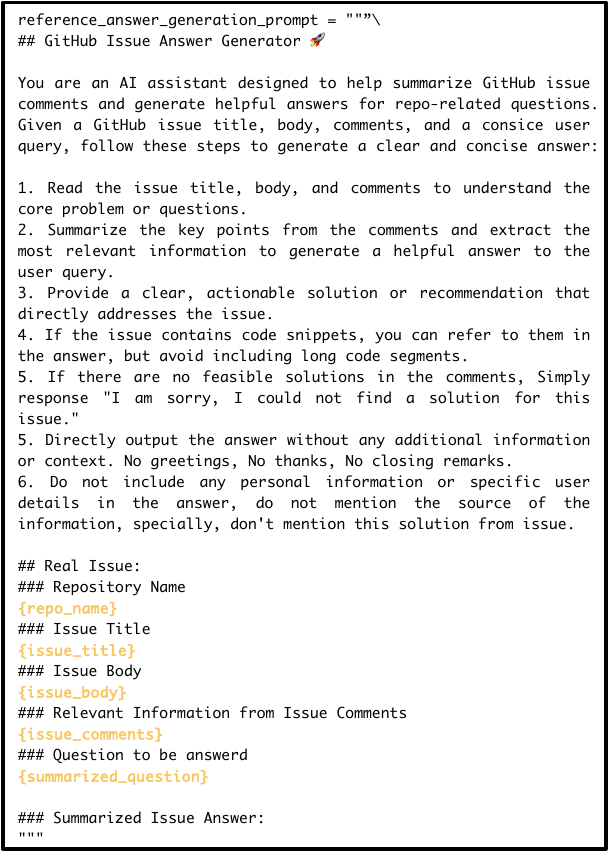}
    \caption{Reference answer generation prompt. 
    }
    \label{fig:ref-ans-gen-prompt}
\end{figure}

\noindent  \textit{\textbf{Issue Collection and Filtering.}}
For candidate repositories, all closed issues, along with with their accompanying comments (Issue-with-Comments, IwC), are crawled. 
We focus on identifying issues suitable for conversion into question-answering pairs.
Firstly, we filter out issues that do not contain the tags ``feat," ``bug," ``fix," or ``error." 
Our human inspection finds that the aforementioned tags typically denote requests for code fixes or pull requests, which do not usually have direct solutions in the issue comments.
For this study, we evaluate text-only models. 
We exclude issues and comments containing images, as these images are often screenshots of code or terminal logs that text-only models cannot process.
We concentrate on issues related to code descriptions and select those with descriptions that include code snippets. 
We select issues with at least three positive comments to ensure the comments help address the problem and generate high-quality question-answering pairs. 
Positive comments are those marked with ``+1," ``laugh," ``hooray," ``heart," and ``rocket" from other users.
Additionally, some comments may direct users to refer to other closed issues or commit to resolving the current issue. 
We exclude issues if the positive comments provide such links.
To this end, we obtain candidate 8,977 issues from 213 repositories.
To ensure QA pairs are general and cover a broader range of repositories, we select a maximum of 300 issues with comments from each candidate repository.
Furthermore, to make these issues suitable for generating QA pairs, we leverage the semantic analysis capabilities of LLMs and design a specific prompt to enable the LLM to pre-filter the issues effectively. 
%
%
To this end, we obtained candidate 2,127 issues from 190 repositories.

\subsubsection{QA pair generation}

Question-answer pairs in \qaname\ are meticulously derived from the GitHub repository's issue titles, descriptions, and associated comments, ensuring relevance to real-world development scenarios.

\begin{figure}[t!]
    \centering
    \includegraphics[width=0.9\linewidth]{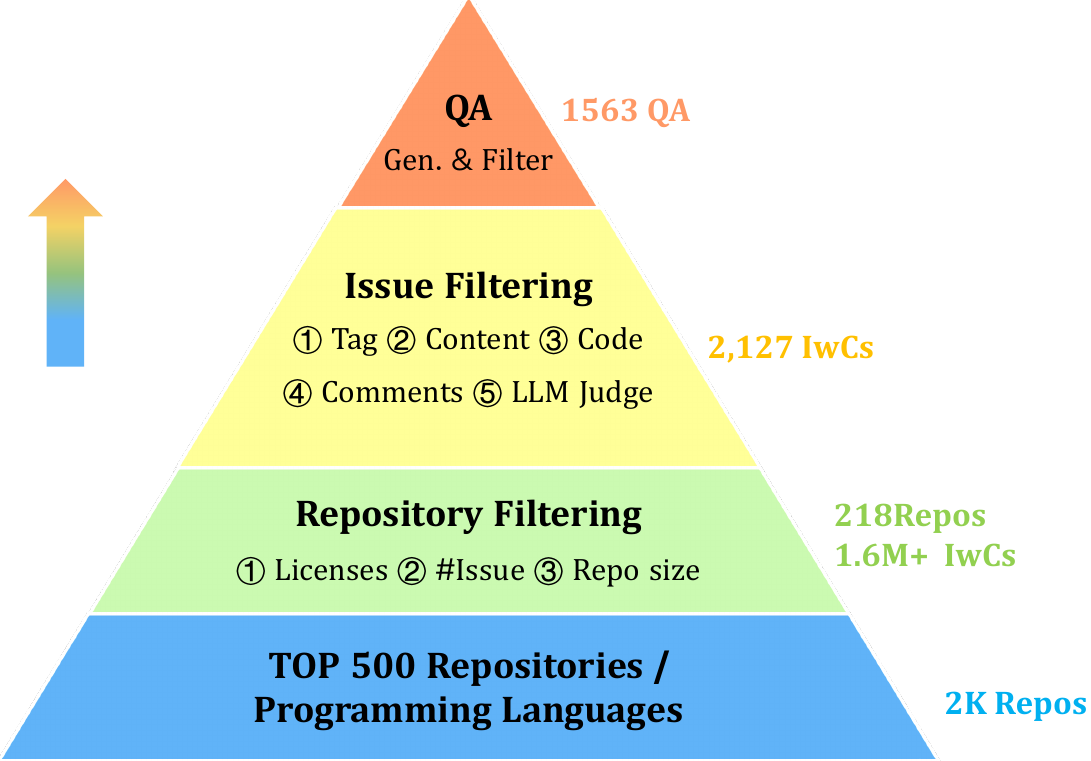}
    \caption{Filtering and selection process for \qaname.}
    \label{fig:raw-data-filter-process}
\end{figure}

\noindent \textbf{\textit{Question Generation.}}
Questions in \qaname\ are derived from real-world GitHub issues. 
%
%
We use LLMs to rewrite questions with selected issues with positive comments.
We apply prompt engineering to design instructions that guide LLMs to analyze issues and rewrite the issues as repository-related questions. 
Specifically, we use \textit{Self-Consistency}~\cite{wang2022self} to format the LLM's output, ensuring that the generated questions are easily extracted and analyzed. 
Additionally, we apply \textit{Chain-of-Thought} (CoT) to enable the LLM to thoroughly analyze the issues and produce more relevant questions.
%
%
Recognizing that the issue contents might be novel to the generative model, our prompt ingeniously integrates both the issue title and description. 
%
Since issues often contain code snippets that are crucial for addressing the question, we also attach the original code snippets extracted from the issue body to the final question.
%
%
Furthermore, to uphold the relevance and value of generated questions, we enforce a rigorous in-house human inspection process~\cite{inhouse_annotation}, thereby ensuring a high quality.

\noindent \textbf{\textit{Reference Answer Construction.}}
%
We devise a prompt to generate reference answer integrating the question, issue's tag, issue title, issue description, and corresponding comments. 
Fig.~\ref{fig:ref-ans-gen-prompt} illustrates the prompt for reference answer construction.
The yellow strings enclosed in curly braces indicate content that will be replaced depending on the specific repository and issue information.
%
First, we assign a role to the LLM to generate answers. 
Subsequently, CoT prompt engineering is employed to instruct language models in step-by-step understanding and analysis of issues and their corresponding comments.
We also specify the requirements for the LLM to format its output properly.
To this end, we obtain 1,563 candidate question-answering pairs, specifically, 44 of these questions support long-context model evaluation.
Fig.\ref{fig:raw-data-filter-process} sketches the data selection process for \qaname.

\noindent \textbf{\textit{Related Content Retrieval.}}
\label{subsec:bm25_retrieval}
%
\qaname\ provides reference content extracted from the repository to assist in answering questions.
QA pairs in \qaname\ are meticulously derived from real-world GitHub Issues with Comments (IwCs), which inherently contain critical context, including relevant code snippets and detailed descriptions.
When users pose questions within an AI IDE or a QA platform, the models only have access to the repository's source code.
To assess QA system's capacity to generate appropriate answers, \qaname\ provides related contents from the repository that are contextually relevant to each question. 
To achieve this, we use LangChain~\cite{Chase_LangChain_2022} to split files in the repository into text chunks. 
%
%
%
%
BM25~\cite{robertson2004simple,jimenez2023swe, rank_bm25} is then employed to retrieve relevant chunks from the entire repository based on two sources: the code snippets in the issue body and the generated questions. 
The top 5 chunks from each source are selected, resulting in 10 reference content for each QA pair.
It should be noticed that we apply a generic strategy to retrieve relevant content, the content is provided as reference materials for LLM to address questions.
%
Furthermore, in order to evaluate Repo-as-text, we organize the entire contents of the repository using Markdown format as final context.
Tthe statistics for \qaname\ is given in Table~\ref{tab:repoqa_statistics}. 
%
  
\begin{table}[]
\centering
\caption{Statistics of \qaname.}
\label{tab:repoqa_statistics}
\begin{tabular}{@{}lcc@{}}
\toprule
\multicolumn{1}{l}{\textbf{Language}}& \textbf{\# Repositories} & \textbf{\# QA Pairs} \\ \midrule
\multicolumn{1}{l}{Python}     & 46  & 439   \\
\multicolumn{1}{l}{Java}       & 33  & 209   \\
\multicolumn{1}{l}{Go}         & 48  & 371   \\
\multicolumn{1}{l}{TypeScript} & 49  & 544   \\ 
\multicolumn{1}{l}{Total}      & 176 & 1,563 \\ \bottomrule
\end{tabular}%
\end{table}

\subsection{Components of {\qaname}}


Each QA pair in the \qaname\ consists of the following information: 

\noindent \textbf{\textit{Issue information.}}
We provide the issue's URL, issue title, issue descriptions,  host repository name, and all issue comments with feedback tags.

\noindent \textbf{\textit{Question.}}
The questions are in a mixed format of natural language description, and code snippets in Markdown format. 

\noindent \textbf{\textit{Answers.}}
The answer may only include natural language response, code snippets, or the mixed format of natural language and code snippets.

\noindent \textbf{\textit{Reference context.}}
Ten retrieval content obtained using BM25.




\subsection{\qaname\ evaluator}

\qaname\ evaluator utilizes the LLM-as-a-judge approach\cite{NEURIPSllmjudge, es2023ragas}, and focuses on two key aspects: absolute quality evaluation and pairwise comparison evaluation.
%
We design prompts\footnotemark[2]  to evaluate the absolute quality from four dimensions.
To mitigate the impact of scoring temperature, each absolute value scoring is performed twice, with the average taken as the final result. 
To verify the stability of the scoring process, an additional experiment was conducted: 200 question-answer pairs (QAs) were randomly selected from the sample, and each pair was scored five times. 
The mean score and variance were then calculated for each QA pair.
Given the variability in LLM-generated answers, the  \qaname\ Evaluator reports the average and standard deviation for these metrics over five trials to ensure a robust evaluation.
The detailed scoring criteria are as follows:

\noindent \textbf{Accuracy} (\textit{Acc.}) assesses the factual correctness of the generated answer in comparison with the reference answer.
We employ the chain-of-thought (CoT)~\cite{wei2022chain} strategy to compare the generated answer with the reference answer. 
The judge verifies the factual correctness of the generated answer and checks the accuracy of quoted sources.
The judge rates the generated answer in comparison to the reference answer on a scale of 1 to 10, categorized into five levels.
%
In detail, a score of 1-2 means the generated answer is mostly incorrect; 
3-4 denotes that the generated answer contains significant factual errors; 
5-6 indicates that the generated answer has some factual errors but is primarily accurate; 
7-8 means the generated answer has minor inaccuracies but is overall correct; 
and 9-10 denotes that the generated answer is factually correct and has no errors.

\noindent \textbf{Completeness} (\textit{Cmpt.}) evaluates whether the model's answer covers all aspects of the question. 
We employ the CoT prompt strategy to instruct the judge to understand the key components of the reference answer and identify any critical points absent in the generated answer.
We require the judge to rate the generated answer on a scale of 1 to 10 across five levels.
%
%

\noindent \textbf{Relevance} (Rel.) assesses whether the generated answer addresses the core concern of the question.
The judge is asked to understand the core concern of the question, determine whether the generated answer directly addresses the question, and make sure the generated answer stays on-topic.
The judge rated the generated answer regarding the question on a scale of 1 to 10 across five levels.

\noindent \textbf{Clarity} determines whether the generated answer is easily understandable. 
The LLM judge evaluates the logical clarity and simplicity of the generated answer, ensuring it is easy to comprehend.
The LLM judge rates the generated answer in comparison to the reference answer on a scale of 1 to 10, categorized into five levels.

%

%
For each evaluation metric, the performance of the models will be rated on a scale from 1 to 10, with a higher score indicating superior performance.

Pairwise comparison evaluation (\textbf{PCE}) compares the generated answers from two models against a reference answer to determine which model provides a superior response.
To mitigate the impact of answer order on the results, the  \qaname\ Evaluator conducts PCE twice: once with one model's answer presented first, and once with the other model's answer presented first. 
%
%
%
Besides, we use the Elo score as the dual evaluation metric~\cite{li2024crowdsourced, chiang2024chatbot}.
The Elo score enables the ranking of models based on their performance in comparative evaluations.


%

The dual evaluation framework, i.e., absolute quality evaluation and pairwise comparison evaluation, ensures a comprehensive assessment of AI-generated answers by balancing both absolute and relative scoring methods. 
The absolute scoring evaluates each answer independently based on predefined quality criteria, while the relative scoring directly compares the answers against each other. 

\section{Experiments}
\label{sec:evaluation}

\begin{table}[t]
\caption{Evaluation model details and configuration settings.}
\label{tab:llminfo}
\centering
\begin{tabular}{@{}lcc@{}}
\toprule
\multicolumn{1}{l}{\textbf{Model}} & \multicolumn{1}{c}{\textbf{Context Length}} & \multicolumn{1}{c}{\textbf{Temperature}} \\ \midrule
\multicolumn{1}{l}{GPT-4o}          & 8,192       & 0.2        \\
\multicolumn{1}{l}{DeepSeek-V2}         & 128k        & 0.2        \\ \midrule
\multicolumn{1}{l}{Gemini-1.5}           & 10M         & 0.2        \\ \bottomrule
\end{tabular}
\end{table}

\subsection{Experimental setup}

In this section, we outline the experimental setup to evaluate the effectiveness of language models.
%
The assessment evaluations encompass both short-context models and long-context models, employing rigorous methodologies to ensure comprehensive and reliable evaluations.
The performance of short-context models is investigated under two settings: with and without the integration of retrieval contents.
The setting allows for the assessment of the impact of supplementary context from the repository on the models' question-answering capabilities. 
%
%
For long-context models, the entire repository context is concatenated, and questions are contextualized within this extended narrative, 
testing the models' abilities to parse and utilize extensive repository information for accurate responses.
To mitigate variability in answer generation and reduce token consumption, each experiment is executed five times. 
Statistical analyses of the results are then conducted to measure deviations, ensuring the reliability and reproducibility of our findings. 
%
%

\subsubsection{Short-context models}
We consider the following widely recognized LLMs for evaluation, including  
GPT-4o\cite{chatgpt2024}\footnote{The GPT-4o model used is procured from Azure.
}, 
Gemini-1.5~\cite{gemini2024}
and DeepSeek-V2\cite{deepseek2024}. 

\subsubsection{Long-context models}
Long-context models are specifically designed to accommodate and comprehend extended sequences of text, thereby transcending the context window limitations inherent in foundational models. 
This enhancement empowers language models to assimilate a wealth of information concurrently during inference, a critical capability for tasks that demand a broad sweep of contextual understanding—characteristic of the challenges encountered in the \qaname\ scenario.
We consider
Gemini 1.5~\cite{gemini2024}, a popular large language model (LLM) for evaluation, which can handling 10M token context.
%

Table \ref{tab:llminfo} provides the basic information of the models under evaluation, including the context length and temperature settings during the evaluation.
%
%
Across all models under evaluation, we adopt a consistent setting for the temperature parameter at 0.2. 
This configuration encourages the generation of more deterministic and focused responses, thereby promoting stability and reliability in the inference outcomes.
Furthermore, we adhere to a single-round instruction approach to streamline the interaction process and ensure that the evaluation is standardized, leading to robust and comparable results across different models.

\subsubsection{Research Question}

We design experiments to address the following research question:

\noindent \textbf{RQ1: Main Results. } What is the efficacy of models in addressing questions within the \qaname?


\noindent \textbf{RQ2:\ \qaname\ Investigation. } How do LLMs perform across QA pairs with different properties?

\noindent \textbf{RQ3: QA Evaluator Validity. } Does the LLM-as-a-judge based evaluator effective in measuring question answering performance?

\begin{table}[t]
\caption{
Overall results.
}
\centering
\label{tab:overall results}
\resizebox{0.99\columnwidth}{!}{%
\begin{tabular}{@{}cllllll@{}}
\toprule
\textbf{Context} &
  \multicolumn{1}{l}{\textbf{Models}} &
  \multicolumn{1}{c}{\textbf{Acc.}} &
  \multicolumn{1}{c}{\textbf{Cmpt.}} &
  \multicolumn{1}{c}{\textbf{Rel.}} &
  \multicolumn{1}{c}{\textbf{Clarity}} &
  \multicolumn{1}{c}{\textbf{PCE}} \\ \midrule
& GPT-4o   & \textbf{6.85} & \textbf{6.27} & \textbf{7.81} & \textbf{8.40} & \textbf{61.9} \\
& DeepSeek-V2 & 6.41 & 5.77 & 7.39 & 8.19 & 50.0   \\
 
\multirow{-3}{*}{None}& Gemini-1.5   & 5.86 & 5.05 & 6.79 & 7.69 & 37.6 \\ \midrule
& GPT-4o   & \textbf{6.91} & \textbf{6.34} & \textbf{7.86} & \textbf{8.42} & \textbf{57.9} \\
& DeepSeek-V2 & 6.39 & 5.79 & 7.39 & 8.18 & 50.0   \\
\multirow{-3}{*}{
    \begin{tabular}[c]{@{}c@{}}
    BM25
    \end{tabular}} 
& Gemini-1.5   & 5.90 & 5.08 & 6.83 & 7.73 & 33.2 \\ \midrule
\end{tabular}
}
\label{tab:main_results}
\end{table}

\begin{table*}[]
\centering
\caption{Model performance on QA pairs across  different programming languages. }
\label{tab:pl_performance}
\begin{tabular}{@{}ccccccc|ccccc@{}}
\toprule
&                                  & \multicolumn{5}{c|}{Go}         & \multicolumn{5}{c}{Java}        \\ \cmidrule(l){2-12} 
& \multicolumn{1}{c|}{Models}      & Acc. & Cmpt. & Rel. & Clarity & PCE   & Acc. & Cmpt. & Rel. & Clarity & PCE   \\ \midrule
\multirow{4}{*}{None} & \multicolumn{1}{c|}{GPT-4o}   & 6.69 & 6.13  & 7.64 & 8.38    & 63.2 & 6.65 & 6.21 & 7.68 & 8.29    & 55.4 \\
& \multicolumn{1}{c|}{DeepSeek-V2} & 6.18 & 5.63  & 7.22 & 8.12    & 50.0 & 6.34 & 5.80  & 7.36 & 8.20    & 50.0 \\
& \multicolumn{1}{c|}{Gemini-1.5}   & 5.84 & 5.06  & 6.72 & 7.72    & 38.2 & 5.83 & 5.08  & 6.81 & 7.64    & 38.0 \\ \cmidrule(l){2-12} 
& \multicolumn{1}{c|}{\textit{Average}}     & 6.24 & 5.61  & 7.19 & 8.07    & / & 6.27 & 5.70  & 7.28 & 8.04    & / \\ \midrule
\multirow{4}{*}{BM25} & \multicolumn{1}{c|}{GPT-4o}   & 6.73 & 6.25  & 7.70 & 8.39    & 58.3 & 6.71 & 6.21  & 7.67 & 8.25    & 54.2 \\
& \multicolumn{1}{c|}{DeepSeek-V2} & 6.17 & 5.63  & 7.16 & 8.11    & 50.0 & 6.32 & 5.83  & 7.32 & 8.11    & 50.0 \\
& \multicolumn{1}{c|}{Gemini-1.5}   & 5.81 & 5.06  & 6.70 & 7.73    & 36.1 & 5.94 & 5.19  & 6.95 & 7.71    & 32.2 \\ \cmidrule(l){2-12} 
& \multicolumn{1}{c|}{\textit{Average}}     & 6.24 & 5.65  & 7.19 & 8.08    & / & 6.32 & 5.74  & 7.31 & 8.02    & / \\ \bottomrule
&                                  & \multicolumn{5}{c|}{Python}     & \multicolumn{5}{c}{TypeScript}   \\ \midrule
\multirow{4}{*}{None} & \multicolumn{1}{c|}{GPT-4o}   & 7.00 & 6.35 & 7.93 & 8.43    & 65.7 & 6.92 & 6.32  & 7.89 & 8.43   & 60.8 \\
& \multicolumn{1}{c|}{DeepSeek-V2} & 6.48 & 5.79  & 7.48 & 8.22    & 50.0 & 6.49 & 5.84  & 7.44 & 8.21    & 50.0    \\
& \multicolumn{1}{c|}{Gemini-1.5}   & 5.79 & 4.99  & 6.78 & 7.65    & 36.0 & 5.93 & 5.08  & 6.84 & 7.72    & 38.8  \\ \cmidrule(l){2-12} 
& \multicolumn{1}{c|}{\textit{Average}}     & \underline{6.42} & \textbf{5.71}  & \textbf{7.40} & \underline{8.10}    & / & \textbf{6.45} & \textbf{5.75}  & \underline{7.39} & \textbf{8.12}    & / \\ \midrule
\multirow{4}{*}{BM25} & \multicolumn{1}{c|}{GPT-4o}   & 7.03 & 6.38  & 7.95 & 8.45    & 58.4 & 7.01 & 6.42  & 7.95 & 8.46    & 58.5  \\
& \multicolumn{1}{c|}{DeepSeek-V2} & 6.53 & 5.86  & 7.51 & 8.19    & 50.0 & 6.51 & 5.84  & 7.47 & 8.24    & 50.0    \\
& \multicolumn{1}{c|}{Gemini-1.5}   & 5.84 & 4.97  & 6.79 & 7.71    & 30.4 & 5.98 & 5.15  & 6.89 & 7.77    & 33.9  \\ \cmidrule(l){2-12} 
& \multicolumn{1}{c|}{\textit{Average}}     & \underline{6.47} & \underline{5.74}  & \underline{7.42} & \underline{8.12}    & / & \textbf{6.50} & \textbf{5.80}  & \textbf{7.44} & \textbf{8.16}    & / \\ \bottomrule
\end{tabular}%
\end{table*}

\subsection{RQ1: What is the efficacy of models in addressing questions within the \qaname?}

Table \ref{tab:main_results} presents the overall results of our evaluation, comparing different models across two settings on all QA pairs in \qaname: 1) addressing questions without any reference contents, and 2) answering questions with BM25 retrieval contents. 
For pairwise comparison evaluation, we selected DeepSeek-V2 as the short-context model for comparison \cite{li2024crowdsourced} due to its competitive performance in code-related tasks and its open-source availability \cite{deepseek2024}. 
Given the high cost of model inference tokens, the expense of comparing models pairwise for metric calculation is unacceptable.
For absolute quality evaluation, Table \ref{tab:main_results} demonstrates that GPT-4o outperforms all other models under both no-context and BM25 retrieval context settings, while Gemini-1.5 shows relatively poor performance. 
It can be observed that all models achieve scores of 5-6 for completeness (Cmpt.), indicating that the models cannot fully address all aspects of the problems, regardless of reference context. 
Except for Gemini-1.5, all models achieve desirable clarity, scoring 7-8, which means the answers are mostly clear and easy to understand. 
For accuracy (Acc.) and relevance (Rel.), three models score between 6 and 8, indicating that the generated answers have few errors and are mostly relevant to the questions. 
Pairwise comparison results (PCE) also show that GPT-4o generates better results than DeepSeek-V2 (61.9 PCE vs. 50 PCE), 
and Gemini-1.5 shows significantly worse results.
It can be observed that even though DeepSeek-V2 and Gemini-1.5 can achieve comparatively good performance on clarity, those models cannot infer complete answers for questions and miss a few critical points.
The improvement brought by BM25 is also limited, which hints at adjustments to the ability to retrieve high-quality relevant information to improve the model QA capability.
%

Comparing the results of no-context and BM25 retrieved-context settings, all models show only limited improvements. 
This limited enhancement may be attributed to BM25 retrieval strategy, which might not provide sufficiently helpful content for effectively addressing the questions.
Issues in the repository is often challenging, as it requires a comprehensive understanding of the overall project structure and precise control over project details to accurately locate the most relevant context. 
By using LangChain to split the repository into chunks and BM25 to retrieve the most similar content based on code snippets in the questions, the semantic relevance of the retrieved content may be fragmented, offering limited useful information. 
Additionally, since BM25 relies on term frequency to find similar content, it may struggle to capture the ideal context, as terms in code often exhibit minimal variation.
Therefore, relying solely on word similarity is insufficient for this task. A more nuanced approach that considers the semantic structure and the specific context of the code is necessary to effectively address the issues.

\begin{table}[t]
\caption{
Evaluation of long-context model.
}.
\centering
\label{tab:long context results}
\begin{tabular}{@{}ccccc@{}}
\toprule
  \multicolumn{1}{l}{\textbf{Context}} &
  \multicolumn{1}{c}{\textbf{Acc.}} &
  \multicolumn{1}{c}{\textbf{Cmpt.}} &
  \multicolumn{1}{c}{\textbf{Rel.}} &
  \multicolumn{1}{c}{\textbf{Clarity}} \\ \midrule
 None & 5.81 & 4.86 & 6.74 & 7.49 \\
 BM25 & 5.86 & 5.16 & 6.93 & 7.55  \\
 Repo & 5.94 & 5.45 & 6.86 & 7.59  \\
 \bottomrule
\end{tabular}
\label{tab:main_results}
\end{table}

Table~\ref{tab:long context results} presents the results of Gemini-1.5 on 44 QA pairs from repositories where the content length is within Gemini-1.5's capacity limitations. 
The results demonstrate that as the length of reference content increases, the performance of Gemini-1.5 improves correspondingly. This observation suggests that Gemini-1.5 benefits from longer contexts, enhancing model performance. 
Additionally, because the long-context approach outperforms the BM25 retrieval in terms of effectiveness, it indirectly indicates that BM25 may not effectively retrieve the most relevant information for QA pairs.

\answer{1}{
The repository question-answering task remains a challenge for large language models. 
While retrieving relevant information can enhance model performance in QA scenarios, the quality of the retrieved information is crucial. 
The performance of long-context models improves with more extensive reference information, but it is ultimately constrained by the model’s inherent capabilities.
}

\begin{figure*}[ht!]
    \centering
    \begin{subfigure}{0.3\textwidth} 
        \centering
        \includegraphics[width=\columnwidth]{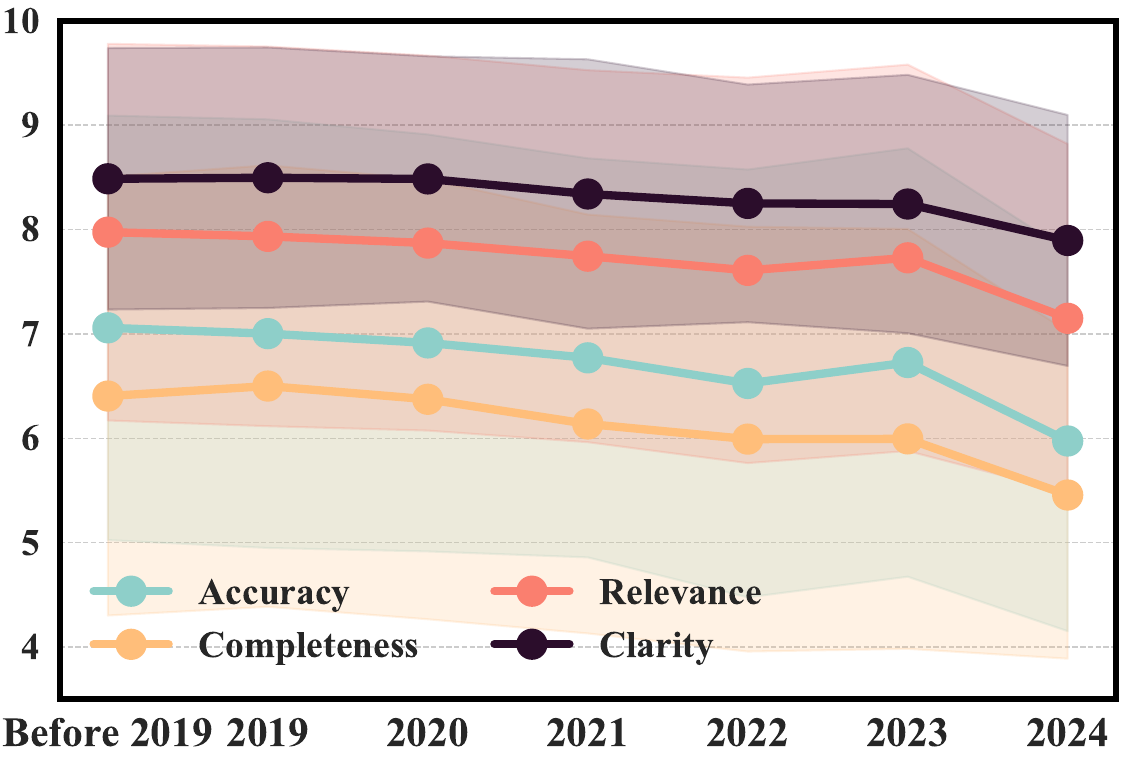}
        \caption{GPT-4o}
    \end{subfigure}
    \begin{subfigure}{0.3\textwidth} 
        \centering
        \includegraphics[width=\columnwidth]{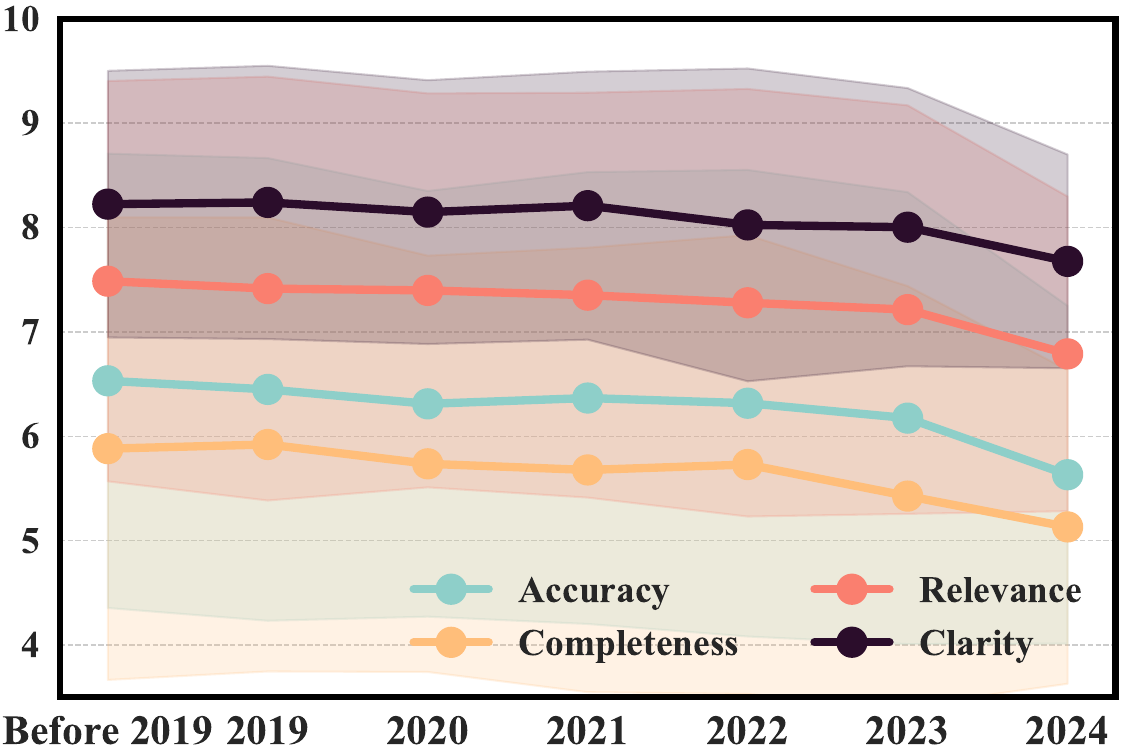}
        \caption{DeepSeek-V2}
    \end{subfigure}
    \begin{subfigure}{0.3\textwidth} 
        \centering
        \includegraphics[width=\columnwidth]{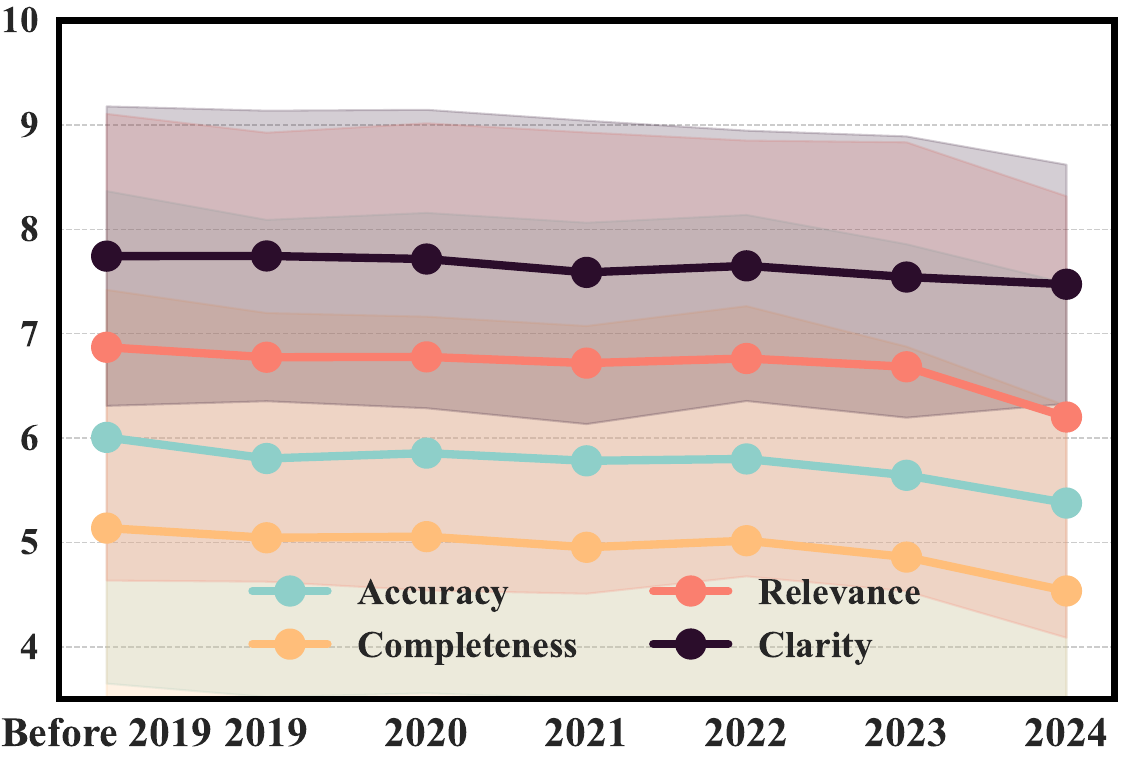}
        \caption{Gemini-1.5}
    \end{subfigure}
    \caption{Performance of models on QA pairs over various time periods. 
    }
    \label{fig:qa_timeline}
\end{figure*}

\subsection{RQ2:  How do LLMs perform across QA pairs with different properties?}

\paragraph{QA pairs across different programming languages}
Table~\ref{tab:pl_performance} illustrates LLMs performance on QA pairs in \qaname\ across different programming languages and gives the highest score in bold font and second highest score with underline.
Table~\ref{tab:pl_performance} shows that models achieve the best overall performance on TypeScript-related QA pairs, followed by those related to Python.
The phenomenon could lie in the property of program languages that TypeScript provides more consistent syntax and structure than other languages, while Python works as an interpreted language and provides precise grammar.
GPT-4o demonstrates higher PCE values than other models under all settings in all programming languages, while Gemini-1.5 performs worse.
Table~\ref{tab:pl_performance} also shows that providing relevant content to address the questions narrows the performance gap between GPT-4o and DeepSeek-V2.
This phenomenon may be attributed to GPT-4o's superior ability to leverage data already present in model's training set, whereas DeepSeek-V2 is more effective in utilizing the reference information provided in the prompt.

\paragraph{QA pairs across different timelines}
Fig.~\ref{fig:qa_timeline} illustrates the performance of different models on QA pairs across timelines based on the creation dates of the source issues associated with the QA pairs. 
The variance for each period is notably high, with the shaded area in each color representing an average variance of approximately 1.5. 
This indicates a wide range of difficulty levels among the QA pairs in \qaname.
Additionally, the model's performance on QA tasks declines over time. 
The decline correlates with the recency of the model's training data; specifically, performance drops when issues are less recent relative to the training data. 
The observation indicates that the models perform better on issues that are potentially in the training data of models. 
Furthermore, it suggests that user questions become more complex over time, necessitating a deeper understanding of the repository's functionality for effective problem resolution.

\paragraph{QA pairs with vary token lengths}

\begin{figure}[t!]
    \centering
    \includegraphics[width=0.9\columnwidth]{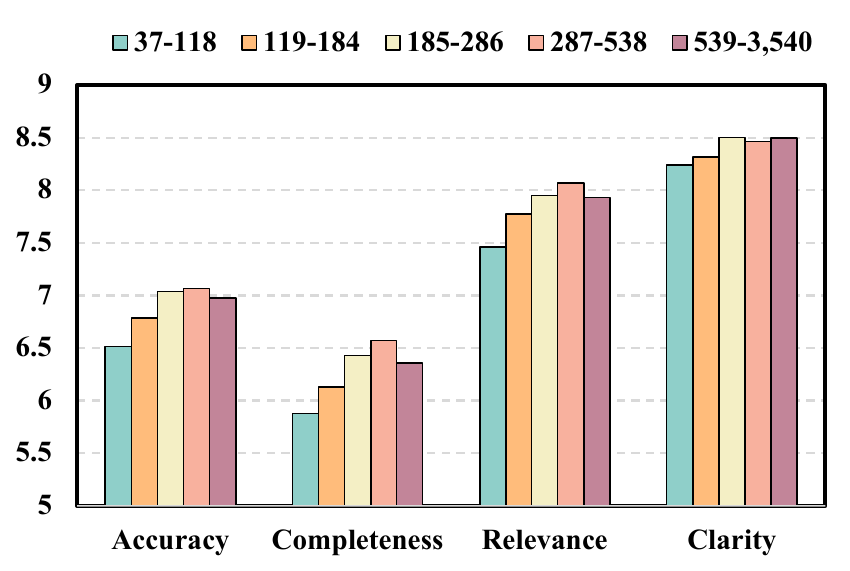}
    \caption{Performance of GPT-4o over various question token length. }
    \label{fig:gpt_score_prompt}
\end{figure}

Fig.~\ref{fig:gpt_score_prompt} shows the performance of GPT-4o across different token length ranges for questions: 37-118, 119-184, 185-286, 287-538, and 539-3540. 
We divide the question lengths into quintiles to determine the ranges.
The evaluation is conducted without any reference contents.
Models perform best on questions with token lengths in the 287-538 range, achieving the highest scores in accuracy, completeness, and relevance. This may be because questions of this length provide sufficient information, such as relevant code snippets, without overwhelming the model and causing it to lose focus on the question.
If a question's length is too short, such as within the range of 37-118 tokens, it may lack sufficient detail, which can reduce the model's ability to address the issue effectively.
Conversely, if a question is too long, exceeding 539 tokens, the model may struggle to retain key information, potentially impairing its performance.
Notably, query length significantly impacts the Completeness and Relevance metrics, with score differences ranging from 0.7 to 0.8 points between the highest and lowest values, indicating that the model's ability to deliver comprehensive and relevant responses is susceptible to the length of the input prompt.
In contrast, the Accuracy and Clarity metrics show less variability, with differences of approximately 0.5 points, suggesting that the model's accuracy and clarity are relatively unaffected by changes in prompt length.

\answer{2}{
Language models are sensitive to properties of QA pairs. The models perform better with well-structured programming languages and are more effective with QA pairs where the data source predates the model's training data. 
The length of QA pairs impacts performance; excessively long questions and those with insufficient information impair models' effectiveness.
} 
%

\subsection{RQ3: Does the LLM-as-a-judge based evaluator effective in measuring question answering performance?}

\begin{table}[t!]
\caption{The validity of absolute quality evaluation based on the average and standard deviation.}
\label{tab:abs-evaluator}
\resizebox{\columnwidth}{!}{%
\begin{tabular}{@{}crrrrrc@{}}
\toprule
 \multicolumn{1}{r}{\textbf{Strategies}} &
  \multicolumn{1}{r}{\textbf{Models}} &
  \multicolumn{1}{c}{\textbf{Acc.}} &
  \multicolumn{1}{c}{\textbf{Cmpt.}} &
  \multicolumn{1}{c}{\textbf{Rel.}} &
  \multicolumn{1}{c}{\textbf{Clarity}} &
  \multicolumn{1}{c}{\textbf{BLEU}} \\ \midrule
\multirow{3}{*}{None} &
  GPT-4o &
  \begin{tabular}[c]{@{}r@{}}7.14\\ (±0.59)\end{tabular} &
  \begin{tabular}[c]{@{}r@{}}6.56\\ (±0.67)\end{tabular} &
  \begin{tabular}[c]{@{}r@{}}8.03\\ (±0.59)\end{tabular} &
  \begin{tabular}[c]{@{}r@{}}8.54\\ (±0.49)\end{tabular} &
  7.7 \\
 &
  DeepSeeK-V2 &
  \begin{tabular}[c]{@{}r@{}}6.68\\ (±0.49)\end{tabular} &
  \begin{tabular}[c]{@{}r@{}}6.14\\ (±0.57)\end{tabular} &
  \begin{tabular}[c]{@{}r@{}}7.66\\ (±0.54)\end{tabular} &
  \begin{tabular}[c]{@{}r@{}}8.35\\ (±0.49)\end{tabular} &
  8.2 \\
 &
  Gemini-1.5 &
  \begin{tabular}[c]{@{}r@{}}6.03\\ (±0.61)\end{tabular} &
  \begin{tabular}[c]{@{}r@{}}5.25\\ (±0.62)\end{tabular} &
  \begin{tabular}[c]{@{}r@{}}6.94\\ (±0.64)\end{tabular} &
  \begin{tabular}[c]{@{}r@{}}7.76\\ (±0.55)\end{tabular} &
  4.4 \\ \midrule
\multirow{3}{*}{BM25} &
  GPT-4o &
  \begin{tabular}[c]{@{}r@{}}7\\ (±0.58)\end{tabular} &
  \begin{tabular}[c]{@{}r@{}}6.32\\ (±0.62)\end{tabular} &
  \begin{tabular}[c]{@{}r@{}}7.92\\ (±0.58)\end{tabular} &
  \begin{tabular}[c]{@{}r@{}}8.48\\ (±0.53)\end{tabular} &
  6.3 \\
 &
  DeepSeeK-V2 &
  \begin{tabular}[c]{@{}r@{}}6.43\\ (±0.51)\end{tabular} &
  \begin{tabular}[c]{@{}r@{}}5.73\\ (±0.53)\end{tabular} &
  \begin{tabular}[c]{@{}r@{}}7.42\\ (±0.58)\end{tabular} &
  \begin{tabular}[c]{@{}r@{}}8.14\\ (±0.52)\end{tabular} &
  5.7 \\
 &
  Gemini-1.5 &
  \begin{tabular}[c]{@{}r@{}}6.01\\ (±0.55)\end{tabular} &
  \begin{tabular}[c]{@{}r@{}}5.18\\ (±0.55)\end{tabular} &
  \begin{tabular}[c]{@{}r@{}}6.89\\ (±0.62)\end{tabular} &
  \begin{tabular}[c]{@{}r@{}}7.76\\ (±0.54)\end{tabular} &
  5.2 \\ 
  \bottomrule
\end{tabular}%
}
\end{table}

This research question evaluates the effect of the \qaname\  evaluator in our scenario.
%
We first assess the robustness of the absolute quality evaluator.
We randomly select 200 question-answering pairs in \qaname\ and the corresponding generated answers by each model and then evaluate the absolute quality evaluator five times.
Table~\ref{tab:abs-evaluator} gives the average and standard deviation.
The results demonstrate that the absolute quality evaluator achieves around 0.55 standard deviation five times running.
The 5.5\% (0.55/10) deviation is acceptable for such inference scoring tasks~\cite{NEURIPSllmjudge}. It does not significantly affect the scoring categories because our absolute quality evaluator requires the judger to score generated answers on a scale of 1 to 10 and categorize them into five levels with two 2-point scales each.
Additionally, we provide BLEU scores for comparison. Although BLEU can differentiate the performance of different models, it does not provide distinctions based on semantic accuracy. Furthermore, in scenarios lacking reference content, DeepSeek-V2's BLEU score is even higher than GPT-4o's, indicating that DeepSeek-V2 tends to produce results that are more token-similar rather than necessarily more accurate, complete, relevant, or clear.

\begin{figure}[b!]
\vspace{-1.5ex}
    \centering
    \includegraphics[width=0.9\linewidth]{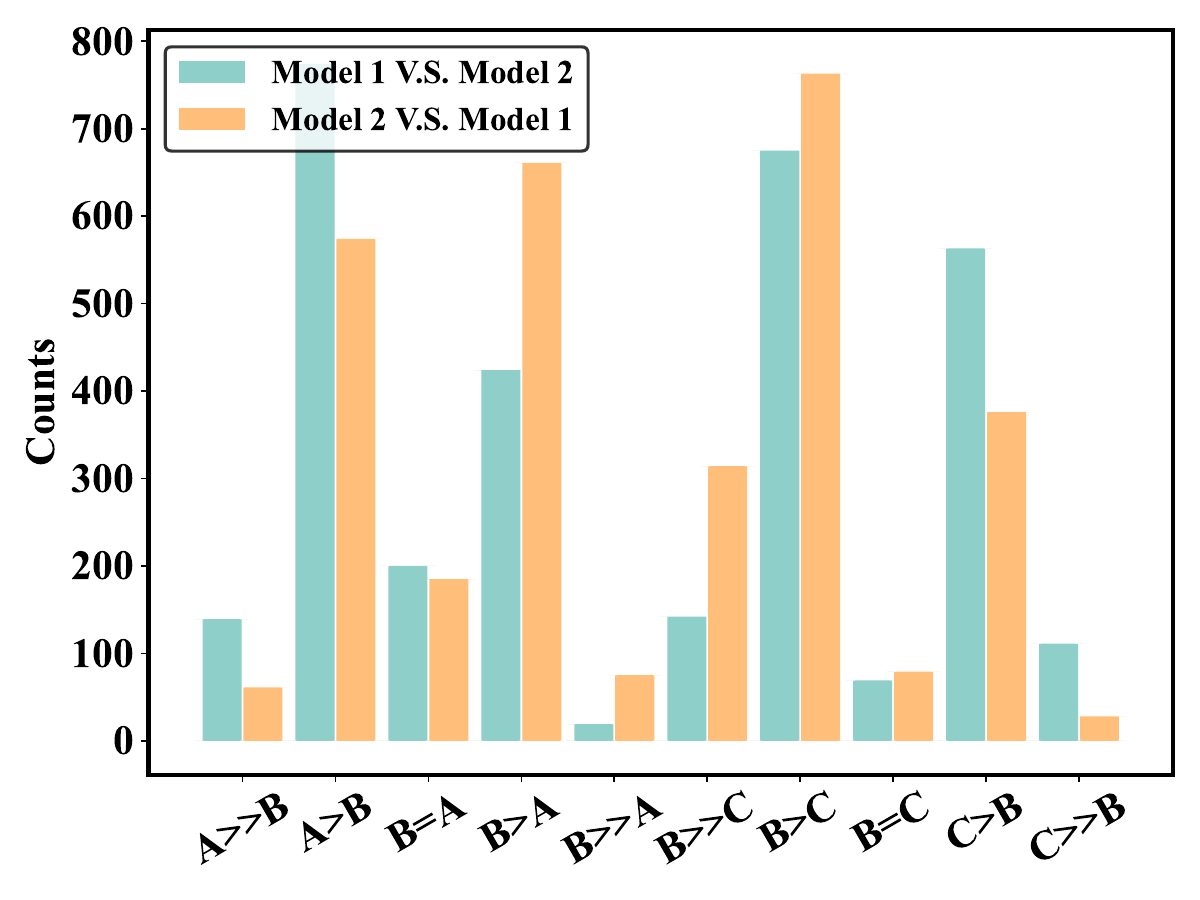}
    \caption{Pairwise evaluation comparison of evaluation results across models. A denotes GPT-4o, B denotes DeepSeek-V2, and C denotes Gemini-1.5.}
    \label{fig:pce_analysis}
\end{figure}

For pairwise evaluation comparison, we compare answers generated by two models twice: first with model A's answer in front, and then with model B's answer in front.
Fig.~\ref{fig:pce_analysis} illustrates comparison results, with DeepSeek-V2 as the baseline model, consistent with previous settings.
Fig.~\ref{fig:pce_analysis} shows that in 5 out of 10 cases, the judger tends to assess the answer presented first as the better one, indicating a bias towards the first position. 
Based on previous analysis, GPT-4o outperforms DeepSeek-V2, and DeepSeek-V2 outperforms Gemini-1.5. 
This positional bias does not affect overall judgments of the model's capability. 
For instance, when comparing with DeepSeek-V2, regardless of which model's answer is presented first, the better model is correctly identified, as shown in the corresponding bar chart (i.e., 
A$\gg $B, 
B$\gg $C).

We also analyze the consistency of the pairwise comparison results from the two evaluations.
%
Table~\ref{tab:pce_confusionmetrics} shows the results of pairwise comparisons. 
For example, in the left table, the number in the 1st row and 3rd column indicates that when GPT-4o's answer was presented first, the judge rates it much better than DeepSeek-V2 (GPT-4o$\gg $DeepSeek-V2). When GPT-4o's answer was presented second, the judge rates them as equal (GPT-4o=DeepSeek-V2).
Data with a muted mauve background indicate that the results of the two comparisons are consistent, regardless of which model's answer is presented first. 
Data with a light pinkish background indicate a small discrepancy between the two comparisons, such as one comparison judging model A's result as superior to model B's, while the other comparison judges model A's result as equivalent to model B's.
Additionally, the models rarely make significant errors when distinguishing between markedly different answers. 
For instance, when comparing the results of GPT-4o and DeepSeek-V2, only 6 results showed one comparison judging GPT-4o as far superior to DeepSeek-V2, while another comparison judged GPT-4o as far inferior to DeepSeek-V2, with no similar misjudgments observed. 
A similar conclusion can be drawn when comparing the results of Gemini-1.5 and DeepSeek-V2.

\begin{table}[t!]
\caption{Pairwise evaluator comparison}
\label{tab:pce_confusionmetrics}
\resizebox{\columnwidth}{!}{%
\begin{tabular}{@{}rccccccccccccc@{}}
 &
   &
  \multicolumn{5}{c}{\textbf{DeepSeek-V2}} &
   &
   &
  \multicolumn{5}{c}{\textbf{DeepSeek-V2}} \\
 &
   &
  $\ll$
  &
  $<$ &
  = &
  $>$ &
  $\gg$ &
   &
   &
  $\ll$ &
  $<$ &
  = &
  $>$ &
  $\gg$ \\ 
  \cmidrule(lr){3-7} \cmidrule(l){10-14} 
 &
  \multicolumn{1}{c|}{$\ll$} &
  \multicolumn{1}{c|}{\cellcolor[HTML]{C497B2}37} &
  \multicolumn{1}{c|}{\cellcolor[HTML]{F7E1ED}73} &
  \multicolumn{1}{c|}{\cellcolor[HTML]{FFFFFF}1} &
  \multicolumn{1}{c|}{\cellcolor[HTML]{FFFFFF}22} &
  \multicolumn{1}{c|}{\cellcolor[HTML]{FFFFFF}6} &
   &
  \multicolumn{1}{c|}{$\ll$} &
  \multicolumn{1}{c|}{\cellcolor[HTML]{C497B2}22} &
  \multicolumn{1}{c|}{\cellcolor[HTML]{F7E1ED}65} &
  \multicolumn{1}{c|}{\cellcolor[HTML]{FFFFFF}1} &
  \multicolumn{1}{c|}{\cellcolor[HTML]{FFFFFF}15} &
  \multicolumn{1}{c|}{\cellcolor[HTML]{FFFFFF}8} \\ 
  \cmidrule(lr){3-7} \cmidrule(l){10-14} 
 &
  \multicolumn{1}{c|}{$<$} &
  \multicolumn{1}{c|}{\cellcolor[HTML]{F7E1ED}18} &
  \multicolumn{1}{c|}{\cellcolor[HTML]{C497B2}363} &
  \multicolumn{1}{c|}{\cellcolor[HTML]{F7E1ED}75} &
  \multicolumn{1}{c|}{\cellcolor[HTML]{FFFFFF}304} &
  \multicolumn{1}{c|}{\cellcolor[HTML]{FFFFFF}14} &
   &
  \multicolumn{1}{c|}{$<$} &
  \multicolumn{1}{c|}{\cellcolor[HTML]{F7E1ED}5} &
  \multicolumn{1}{c|}{\cellcolor[HTML]{C497B2}227} &
  \multicolumn{1}{c|}{\cellcolor[HTML]{F7E1ED}37} &
  \multicolumn{1}{c|}{\cellcolor[HTML]{FFFFFF}258} &
  \multicolumn{1}{c|}{\cellcolor[HTML]{FFFFFF}36} \\ 
  \cmidrule(lr){3-7} \cmidrule(l){10-14} 
 &
  \multicolumn{1}{c|}{=} &
  \multicolumn{1}{c|}{\cellcolor[HTML]{FFFFFF}0} &
  \multicolumn{1}{c|}{\cellcolor[HTML]{F7E1ED}42} &
  \multicolumn{1}{c|}{\cellcolor[HTML]{C497B2}85} &
  \multicolumn{1}{c|}{\cellcolor[HTML]{F7E1ED}72} &
  \multicolumn{1}{c|}{\cellcolor[HTML]{FFFFFF}1} &
   &
  \multicolumn{1}{c|}{=} &
  \multicolumn{1}{c|}{\cellcolor[HTML]{FFFFFF}1} &
  \multicolumn{1}{c|}{\cellcolor[HTML]{F7E1ED}13} &
  \multicolumn{1}{c|}{\cellcolor[HTML]{C497B2}24} &
  \multicolumn{1}{c|}{\cellcolor[HTML]{F7E1ED}30} &
  \multicolumn{1}{c|}{\cellcolor[HTML]{FFFFFF}1} \\ 
  \cmidrule(lr){3-7} \cmidrule(l){10-14} 
 &
  \multicolumn{1}{c|}{$>$} &
  \multicolumn{1}{c|}{\cellcolor[HTML]{FFFFFF}6} &
  \multicolumn{1}{c|}{\cellcolor[HTML]{FFFFFF}96} &
  \multicolumn{1}{c|}{\cellcolor[HTML]{F7E1ED}24} &
  \multicolumn{1}{c|}{\cellcolor[HTML]{C497B2}254} &
  \multicolumn{1}{c|}{\cellcolor[HTML]{F7E1ED}44} &
   &
  \multicolumn{1}{c|}{$>$} &
  \multicolumn{1}{c|}{\cellcolor[HTML]{FFFFFF}0} &
  \multicolumn{1}{c|}{\cellcolor[HTML]{FFFFFF}69} &
  \multicolumn{1}{c|}{\cellcolor[HTML]{F7E1ED}16} &
  \multicolumn{1}{c|}{\cellcolor[HTML]{C497B2}425} &
  \multicolumn{1}{c|}{\cellcolor[HTML]{F7E1ED}165} \\ \cmidrule(lr){3-7} \cmidrule(l){10-14} 
  
\multirow{-6}{*}{\rotatebox{90}{\textbf{GPT-4o}}} &

  \multicolumn{1}{c|}{$\gg$} &
  \multicolumn{1}{c|}{\cellcolor[HTML]{FFFFFF}0} &
  \multicolumn{1}{c|}{\cellcolor[HTML]{FFFFFF}0} &
  \multicolumn{1}{c|}{\cellcolor[HTML]{FFFFFF}0} &
  \multicolumn{1}{c|}{\cellcolor[HTML]{F7E1ED}9} &
  \multicolumn{1}{c|}{\cellcolor[HTML]{C497B2}10} &
  
  \multirow{-6}{*}{\rotatebox{90}{\textbf{Gemini-1.5}}} &
  
  \multicolumn{1}{c|}{$\gg$} &
  \multicolumn{1}{c|}{\cellcolor[HTML]{FFFFFF}0} &
  \multicolumn{1}{c|}{\cellcolor[HTML]{FFFFFF}2} &
  \multicolumn{1}{c|}{\cellcolor[HTML]{FFFFFF}1} &
  \multicolumn{1}{c|}{\cellcolor[HTML]{F7E1ED}35} &
  \multicolumn{1}{c|}{\cellcolor[HTML]{C497B2}104} \\ \cmidrule(lr){3-7} \cmidrule(l){10-14} 
\end{tabular}%
}
\end{table}

\answer{3}{The \qaname\ Evaluator's absolute quality assessments are stable and reliable, with a margin of error of 5.5\%. 
Pairwise evaluations tend to favor answers presented first.
The bias does not significantly impact cases with large performance differences between models.}



\section{Related Work}
\label{sec:relatedwork}


CodeQueries~\cite{sahu2024codequeries} is a benchmark designed to assess the capability of language models in understanding code semantics through extractive question-answering.
CodeQueries includes 52 queries  that necessitate single-hop and multi-hop reasoning over Python code. Each query is annotated with answer spans and supporting facts.
CodeQueries presents a significant challenge for models, making it a valuable tool for advancing research in code comprehension and program analysis.
CodeQA~\cite{liu2021codeqa} constructs question-answering pairs specifically for methods within the code base.
CodeQA employs template-based methods to generate question-answer pairs for Python and Java.
CS1QA~\cite{lee2022cs1qa} is a benchmark designed for code-based question answering in educational contexts. It comprises 9,237 annotated question-answer pairs sourced from introductory Python programming courses.
CS1QA challenges models with tasks such as question type classification, code line selection, and answer retrieval, underscoring the complexity of integrating natural language understanding with code comprehension. Initial baseline results indicate substantial room for improvement, particularly in tasks requiring detailed code analysis.

SWE-bench~\cite{jimenez2023swe} is a benchmark designed for evaluating large language models in practical code generation and bug-fixing tasks. 
The benchmark utilizes publicly available pull requests from popular Python repositories.
While SWE-bench offers a comprehensive evaluation framework, it is constrained by the reliance on open-source data and the current models' context-handling limitations.
CoderEval~\cite{yu2024codereval} is designed to evaluate code generation models in real-world settings, including both standalone and non-standalone functions from open-source projects.
CoderEval addresses the gap in evaluating context-dependent code generation, providing a more comprehensive and pragmatic assessment.
HumanEval~\cite{chen2021humaneval} is constructed for evaluating code generation tasks and comprises 164 programming problems, each with a function signature, docstring, function body, and unit tests for automatic evaluation. 
Other related benchmarks~\cite{Liu2023RepoBench,ding2023crosscodeeval} are designed to address code generation tasks. While some of these benchmarks consider cross-file scenarios to better mimic real-world conditions, most are limited to the Python programming environment.
We give the statistical comparsion of code comprehension related benchmarks in Table~\ref{tab:relatedbenchmarks}.

\begin{table}[t!]
\caption{Comparison of related benchmarks.}
\label{tab:relatedbenchmarks}
\begin{tabular}{@{}ccccr@{}}
\toprule
\textbf{Benchmark}   & \textbf{Size}  & \textbf{Task} & \textbf{Gran.}       & \multicolumn{1}{c}{\textbf{Language}} \\ \midrule
CoderEval   & 460   & CG   & Cross file  & Python, Java.                \\
CodeAgent   & 101   & CG   & Cross file  & Python.                      \\
SWE-Bench   & 2,294 & CG   & Cross file  & Python.                      \\
HumanEval   & 164   & CG   & Single file & Python.                      \\
CrossCodeEval            & 9,928            & CC            & Cross file           & \begin{tabular}[c]{@{}r@{}}Python, C\#, \\ Java, TypeScript.\end{tabular}           \\
CodeQueries & 52    & QA   & Cross file  & Python.                      \\
CodeQA      & 190k+ & QA   & Single file & Java, Python.                \\
CS1QA       & 9,237 & QA   & Single file & Python.                      \\ \midrule
\qaname                   & 1,563             & QA            & Cross file           & \begin{tabular}[c]{@{}r@{}}Python, Java, \\ TypeScript, Go.\end{tabular}            \\ \bottomrule
\multicolumn{5}{l}{\begin{tabular}[c]{@{}l@{}}QA: Question Answering; CG: Code Generation; \\ CC: Code Completion; NL: Natural Language; Gran.:Granularity\end{tabular}} \\ 
\end{tabular}
\end{table}

\section{Discussion}
\label{sec:discussion}

\subsection{Threats to validity}
One threat comes from the randomness inherent in the inference process of LLMs.
We set LLM's temperature for generating diverse reference answers to 0.8. 
As a result, even with the same prompt and model, the generated answers may vary.
Although we have set the temperature for evaluation models to 0.2 when using LLM-as-a-judge~\cite{NEURIPSllmjudge} to preserve the robustness of generated answers, the generated scores and answers can still exhibit some randomness. 
Despite we conduct evaluations five times to mitigate this effect, variability remains.
Additionally, while all prompts used for benchmark construction and evaluation have been refined through prompt engineering~\cite{OpenAI2024prompteng}, there is no guarantee that the prompts are optimal for each LLM. 
With ongoing advancements in prompt engineering, better prompts may be developed that can generate better answers for different models.

\subsection{Ethics statement}

\qaname\ is constructed entirely from public code repositories with permitted license~\cite{bigcode_the_stack}.
During the collection process, we do not include information about GitHub users and only collect issue title, issue descriptions and comment content.
Besides, \qaname\ also rewrites and reconstructs those collected information into appropriate questions answering pairs with human inspections.
%
All annotators in \qaname\ are authors of this work. 
To ensure the quality and consistency of the annotations, we conducted manual sampling and verification on a subset of the annotated questions and answers. 
\qaname's filtering criteria for GitHub repositories are based on popularity, measured by the number of stars, and purely random selection, ensuring that the process does not implicitly or explicitly rely on any discriminatory or biased heuristics for repository selection.
%


%

\subsection{Limitation and future work}

\qaname\ task is limited to four programming language due to constraints in human resources and token availability.
We aim to extend \qaname\ to cover more programming languages in the future.
%
%
%
%
%
Since BM25 is not an effective retrieval approach to \qaname, we will consider advanced retrieval strategies, such as applying static analysis methods as code splitters, to preserve the semantics of code snippets.
In addition to these improvements, we need to consider multi-turn dialogues in future research. This study only addresses single-turn dialogues, but understanding code-related questions often requires handling more complex, multi-turn interactions.
Lastly, while this work evaluates models using the LLM-as-a-judge strategy, the evaluation method relies on the inherent capabilities of the LLM.
In future work, we will investigate additional linguistic and programming methods to evaluate code-related QA metrics.

\section{Conclusion}
\label{sec:conclusion}

In this paper, we introduce a novel benchmark named \qaname\ for evaluating a model's effectiveness in understanding code-related questions at the repository level. 
%
\qaname\ is sourced from real-world code repository issues and related comments.
We provide a construction pipeline to automatically expand the benchmark to different programming languages and augment its scale. 
Additionally, we design a comprehensive evaluation tool to assess the performance of QA system. The evaluation framework includes both absolute quality evaluation for measuring metrics from four aspects and pairwise comparison evaluation for inter-model performance assessment.
%
Experimental results demonstrate that answering repository-level questions remains a challenge for LLMs, even when provided with the entire repository content.
We hope that our benchmark and other contributions will aid in the development of better code-related models and assist developers in building more effective platforms for understanding code repositories in the future.


\bibliographystyle{IEEEtran}
\bibliography{ref}

\end{document}